\documentclass[]{article}
\usepackage{amsmath}
\usepackage{graphicx}
\usepackage{amssymb}
\usepackage{dcolumn}
\usepackage{bm}
\usepackage{amsmath}
\usepackage[utf8]{inputenc}
\usepackage[margin=1.0in]{geometry}
\usepackage{float}
\usepackage[T1]{fontenc} 
\usepackage[linktoc=all]{hyperref}
\hypersetup{
		colorlinks=true,
		linkcolor=blue,
		citecolor=red}
	
\title{Bulk viscous late acceleration under near equilibrium conditions in \boldmath $f(R,T)$ gravity with mixed dark matter.}

\author{Vishnu A Pai$^{*}$ and Titus K Mathew$^{\dagger}$ \\ \small \textit{Department of Physics, CUSAT, Kalamasserry, Kochi 682022, Kerala, India} \\ \small vishnuajithj@gmail.com$^{*}$ and titus@cusat.ac.in$^{\dagger}$ }

\begin{document}
	\maketitle
%	\flushbottom
\begin{abstract}
	Various studies have shown that the late acceleration of the universe can be caused by the bulk viscosity associated with dark matter. But recently, it was indicated that a cosmological constant is essential for maintaining Near Equilibrium Conditions (NEC) for the bulk viscous matter during the accelerated expansion of the universe. In the present study, we investigate a model of the universe composed of mixed dark matter components, with viscous dark matter (vDM), and inviscid cold dark matter (CDM) as it's constituents, in the context of $f(R,T)$ gravity and showed that the model predicts late acceleration by satisfying NEC throughout the evolution, without cosmological constant. We have also compared the model predictions with combined Type Ia Supernovae and observational Hubble data sets and thereby determined the estimated values of different cosmological parameters.
\end{abstract}


\section{Introduction} \label{section 1}

	Explaining late-accelerated expansion of the universe using bulk viscous property of the matter sector is an intriguing possibility. Such cosmological models are special in the sense that, they are devoid of any exotic dark energy component and the negative pressure necessary for the cosmic acceleration is generated via bulk viscosity. Theoretically, inclusion of dissipative effects in matter sector seems to be obvious as the concept of an ideal fluid used in the concordance model, could only be an approximation to reality. Hence, by considering dissipative traits for the fluid, one can gain better understanding about the evolution of the universe. From previous studies in inflationary cosmology \cite{RMaartens_1995, PhysRevD.33.1839}, it was already established (well before the SNe Ia observations) that, fluids with bulk viscosity has the innate ability to predict an accelerated expansion of the universe, that too in the absence of cosmological constant or any exotic cosmic component. This is the fact that motivates authors to investigate the possibilities of viscous driven late acceleration of the universe. 
	
	For incorporating dissipative effects in cosmology, there exists different formalism's in the literature, which are proposed on the basis of relativistic hydrodynamics. Widely investigated class of dissipative models are based on Eckart's and Landau-Lifshitz formalism \cite{PhysRev.58.919, landau2013fluid}. Despite of it's drawbacks, such as the acausal nature of the resulting solutions and unstable behavior of the equilibrium states, Eckart's formalism is considered to be a good first order approximation for investigating viscous nature in the cosmological context. More general formalism, such as Israel-Stewart theory or its truncated version, \cite{1967ZPhy..198..329M, Israel:1976tn, ISRAEL1976213, Israel:1979wp}, which includes higher order corrections from equilibrium, are also used to discuss the dissipative evolution of the universe. Additionally, in recent studies, a new approach for incorporating dissipative effects in comic matter, have also been proposed \cite{PhysRevE.56.6620, PhysRevD.60.103507, PhysRevD.61.023510, OTTINGER1998433, PhysRevD.98.104064, PhysRevD.100.104020, Bemfica:2020zjp, Kovtun:2019hdm, Hoult:2020eho}. Nevertheless, owing to it's simplicity, we follow Eckart theory in the present work to study of viscous cosmology, since the higher order theories pose difficulties in cosmological modeling. 
	
	All relativistic dissipative theories, from acausal Eckart's theory to full causal IS theory, are developed under the assumption that the viscous fluid remains in a near-equilibrium state. Hence, such theories can at most allow only a minute deviation from the local equilibrium pressure of the viscous fluid. However, in \cite{RMaartens_1995}, Marteens inferred that, for explaining the inflation as driven by bulk viscosity of matter, the near equilibrium condition has to be violated. Hence, the author proposed that, in cases where theories with dissipative phenomena are used to explain the accelerated expansion, one is forced to postulate the validity of such theories even in regimes where fluid is far-from equilibrium. Following this assumption, several cosmological models have been proposed to explain late acceleration of the universe
	\cite{mohan2017bulk, sasidharan2015bulk, avelino2009can, avelino2010exploring, brevik2017viscous}, which are, reasonably successful in predicting the cosmological evolution. But the postulate regarding the validity of the dissipative phenomena in situations where fluid is far from equilibrium, still remains only as an assumption which has no definite proof at al. Hence, it turns out that, the safest way to explain the accelerated expansion in the presence of viscous matter, is by maintaining the validity of near equilibrium condition (NEC) throughout the evolution. In technical terms, NEC suggests that the magnitude of viscous pressure of the fluid ($\Pi$) should not exceed the magnitude of its equilibrium pressure ($p^0$), i.e.,
	\begin{equation} \label{1}
		\frac{\big |\Pi\big |}{p^0} \ll 1.
	\end{equation}
	This immediately rules out the possibility of explaining cosmic acceleration, by associating bulk viscosity to the cold dark matter (CDM) component, since the kinetic pressure of it is zero and hence it inevitably violates NEC.
	
	In recent studies on late acceleration by Cruz et al. \cite{Cruz:2022zxe,sym14091866,PhysRevD.105.024047}, based on Eckart's theory, it was inferred that the NEC can be satisfied for a particular viscous model ($\zeta = \zeta_{0} \rho$) in the context of Einstein's gravity, with warm dark matter (WDM) in the presence of a cosmological constant. This implies that, the presence of cosmological constant is deemed inevitable for the validity of the NEC in the context of Einstein gravity. It turns out that the presence of cosmological constant can decrease the effect of viscous pressure $\Pi,$ in generating late acceleration, so that $|\Pi/p^0| \ll 1.$ Since these works demands the presence of cosmological constant for the validity of NEC in the context of Einstein's gravity	it is of great significance to investigate dissipative models of recent acceleration by satisfying the NEC without cosmological constant, in the context of modified gravity theories. In the present work, we are analyzing such a possibility in the context of $f(R,T)$ gravity theory.
	
	Modified gravity theories are of much importance for obtaining suitable description of gravity at large scales, relevant to study of the universe \cite{doi:10.1142/S0218271802002025, PADMANABHAN2013115, PhysRevD.75.084031, PhysRevD.79.084008, PhysRevD.60.043501, nojiri2005gauss, DVALI2000208, capozziello2003curvature, allemandi2004accelerated, nojiri2004modified, dolgov2003can, nojiri2006modified, carroll2004cosmic, nojiri2011unified, sotiriou2010f, de2010f,2010EPJC...70..373H, galaxies2030410}. The $f(R,T)$ gravity \cite{harko2011f}, is a generalization of Einstein's gravity, where the gravitational Lagrangian is assumed to be an arbitrary function of the Ricci scalar, $R$ and the trace of the energy momentum tensor of the cosmic component, $T$. This assumed functional form implies a minimal/non-minimal coupling between geometry of the spacetime and matter. Such a coupling between geometry and the trace of energy-momentum tensor can be induced by the imperfect nature of the fluid or through quantum effects. One of the ways in which trace of matter stress energy tensor could enter the gravitational Lagrangian is through the quantum effect called Trace anomaly or conformal symmetry breaking \cite{Xu:2016rdf,galaxies2030410}. Modified gravity models with geometry-matter coupling are significant since they can also provide a theoretical explanation for the late-time acceleration of the Universe, without postulating the existence of dark energy. In addition, such theories are also known to provide an explicit breaking of equivalence principle which is constrained using solar system experiments \cite{galaxies2030410, Faraoni:2004pi, Bertolami:2006js, PhysRevD.90.044031}. One of the effect of this coupling is the appearance of extra terms in the conservation law for matter component, which has the effect of an orthogonal force on test particles which in-turn renders their motion to be non-geodesic \cite{harko2011f}. In \cite{PhysRevD.90.044067}, authors interpret the appearance of such extra terms, as due to particle creation process. However, Azevedo and Avelino \cite{PhysRevD.99.064027} strongly argued that, such extra terms are not really implying particle creation process, rather, they are to be taken as effective contribution to particles momentum, in cosmological time scales. Authors also point out that, there occurs an active energy exchange between matter and spacetime (without changing the number of particles), due to which the evolution process becomes thermodynamically non-adiabatic. In the present work, we follow this new approach where the extra terms are considered as non-adiabatic contributions to particle four momentum. 
	
	It is to be noted that, almost all previous studies of viscous models in the context of  $f(R,T)$ gravity were done without respecting the NEC \cite{Koussour:2021flk, baffou2015cosmological, Prasad:2020qug, singh2014friedmann, Arora:2020xbn,Debnath:2020bno,Yadav:2020vih}. Our primary aim is to construct a viable viscous cosmological model in the context of $f(R,T)$ gravity that explains the late accelerated expansion of the universe by satisfying NEC, without the cosmological constant. For this, we consider a model of the universe with mixed dark matter components. The choice of mixed dark matter is due to the fact that the $f(R,T)$ model with only viscous dark matter, can satisfy NEC, but lead to an ever-accelerating phase, and thus fail to predict the observed transition \cite{2023arXiv230314451P}. We also point out that, considering a universe with mixture of dark matter components is not at all new in cosmology \cite{cheung2014simplified, boyarsky2009lyman, kamada2016constraints, schneider2018constraining, harada2016structure}. However, different from previous works, here we assume viscous dark matter (vDM - which can be stiff/hot/warm) and inviscid CDM as the constituents of mixed dark matter component. The vDM component has both kinetic and bulk viscous pressures, while the inviscid CDM component remains pressure-less throughout the evolution. This implies a significant difference in the definition of their energy momentum tensors and hence their traces. Owing to this difference, each component can have different explicit coupling with geometry. 
	
	In the present work, we propose a modified Lagrangian by accommodating the traces corresponding to the vDM and CDM, and then formulate the effective field equation. From the field equation, the Friedmann equations and continuity equation for the two cosmic components are obtained by assuming FLRW metric for the spacetime. We will then develop the general constraints on the model parameters, that are necessary for satisfying (i) NEC, (ii) critical energy condition (CEC) and (iii) the second law of thermodynamics (SLT). Analytical solution for the Hubble parameter is then derived by assuming a phenomenological form of coefficient of bulk viscosity, as $\zeta=\zeta_1 H^{-1} \rho + \zeta_0 H$ \cite{Gomez:2022qcu}, where $\zeta_1$, and $\zeta_0$ are constant parameters. We will show that the cosmological predictions in this model depends strongly on the value of coupling parameter $\tilde{\lambda},$ between matter and geometry of spacetime. For a specific choice of $\tilde{\lambda}$ the model is capable of showing $\Lambda$CDM like behavior. We also investigate the special case with $\zeta_{0}=0$ because of its interesting dynamical behavior. Finally, we will compare these models with the combined observational Hubble data (OHD) and Type Ia supernovae (SNe Ia) data, to extract value of model parameters and test feasibility of each model.
	
\section{Field equations in $f(R,T)$ gravity with mixed dark matter components} \label{section 2}

Motivated by conformal anomaly in quantum mechanics, one can generalize $f(R)$ gravity by assuming a trace dependent minimal/non-minimal coupling in the gravitational Lagrangian and hence consider the possibility of $f(R,T)$ gravity \cite{harko2011f}. The most general action in $f(R,T)$ gravity can then be expressed as,
\begin{equation}\label{2}
	S = \frac{1}{16\pi}\int f(R,T)\sqrt{-g}\;d^{4}x + \int L_{m}\sqrt{-g}\;d^{4}x.
\end{equation} 
where $L_{m}$ is the matter Lagrangian and $g=|g_{\mu \nu}|$. Throughout the analysis we have followed (+,-,-,-) signature for the metric tensor and have set $G=c=1$. In the present case, we have two dark matter components, CDM and vDM, having two different energy momentum tensors $T^{(m)}_{\mu \nu}$ and $T^{(vm)}_{\mu \nu}$. Obviously, one can also expect the trace of these energy momentum tensors (i.e, $T_m$ and $T_{vm}$) to differ from each other. Owing to this reason, their coupling with geometry of the spacetime can be different, and to incorporate this change, we modify the above action as, 
\begin{equation}\label{3}
	S = \frac{1}{16\pi}\int f(R,T_{m} ,T_{vm})\sqrt{-g} \;d^{4}x + \int L_{m}\sqrt{-g}\;d^{4}x + \int L_{vm}\sqrt{-g}\;d^{4}x.
\end{equation}
Here, $f(R,T_{m} ,T_{vm})$ represents some arbitrary function of Ricci scalar and traces of both components, and $ L_{m}, \, \, \& \, L_{vm}$ are the Lagrangian of CDM and vDM respectively. These Lagrangians satisfy the relation,
\begin{equation} \label{4}
	T^{(i)}_{\mu \nu} = -\frac{2}{\sqrt{-g}} \frac{\delta \left(\sqrt{-g} \; L_{i}\right)}{\delta g^{\mu \nu}},
\end{equation}
where, $i$ denotes the cosmic component that is involved. For instance, $i=m$ for CDM component and $i=vm$ for the vDM. By assuming that the matter Lagrangian densities, $L_m$ and $L_{vm}$ depends only on metric tensor components \cite{harko2011f}, we vary the action (\ref{2})  with respect to metric $g^{\mu \nu}$ and obtain the equation,
\begin{multline}\label{5}
	\delta S = \frac{1}{16\pi} \int \Big[  f_{T_{m}} \frac{\delta T_{m} }{\delta g^{\mu \nu}} \delta g^{\mu \nu} + f_{T_{vm}} \frac{\delta T_{vm}}{\delta g^{\mu \nu}}\delta g^{\mu \nu} + 16\pi \frac{1}{\sqrt{-g}} \frac{\delta \left(\sqrt{-g}\;L_{m}\right)}{\delta g^{\mu \nu}}  \\ \left. + 16\pi \frac{1}{\sqrt{-g}} \frac{\delta \left(\sqrt{-g}\;L_{vm}\right)}{\delta g^{\mu \nu}}+ f_{R}\delta R - \frac{g_{\mu \nu} f}{2}  \delta g^{\mu \nu} \right] \sqrt{-g} \; d^{4}x.
\end{multline}
Here, we denoted, $f = f\left(R,T_{m} ,T_{vm}\right)$, $f_{T_i}= \partial f/ \partial T_i$ and $f_{R}= \partial f/ \partial R$. The variation of Ricci scalar is then given by,
\begin{equation}\label{6}
	\delta R = R_{\mu \nu} \delta g^{\mu \nu} + g^{\mu \nu} \box \delta g^{\mu \nu} - \nabla_{\mu} \nabla_{\nu} \delta g^{\mu \nu}
\end{equation}
The derivative of the trace of energy-momentum tensors with respect to metric components can be expressed as,  
\begin{equation}\label{7}
	\frac{\delta T_{i} }{\delta g^{\mu \nu}}	=\frac{\delta \left(g^{\alpha \beta} T^{(i)}_{\alpha \beta}\right)}{\delta g^{\mu \nu}} = T^{(i)}_{\mu \nu} + \Theta^{(i)}_{\mu\nu}.
\end{equation}
Where $\Theta^{(i)}_{\mu\nu}$ is given by,
\begin{equation}\label{8}
	\Theta^{(i)}_{\mu\nu} \equiv g^{\alpha \beta} \frac{\delta T^{(i)}_{\alpha \beta}}{\delta g^{\mu \nu}}.
\end{equation}
Once we know the energy momentum tensors and the Lagrangian densities of the assumed components, it is possible to express $\Theta^{(i)}_{\mu\nu}$ of that component as,
\begin{equation}\label{9}
	\Theta^{(i)}_{\mu\nu}=-2T^{(i)}_{\mu\nu}+g_{\mu\nu}L_{(i)}-2g^{\alpha \beta}\frac{\partial^{2}L_{(i)}}{\partial g^{\mu\nu}\partial g^{\alpha \beta}},
\end{equation}
in which $\alpha$ and $\beta$ are summing indices.  By extremizing (\ref{5}) we get the field equation for $f(R,T_m,T_{vm})$ gravity as,
\begin{equation}\label{10}
	f_{R} R_{\mu \nu} - \frac{1}{2}f g_{\mu \nu}  + \left( g_{\mu \nu}\Box -  \nabla_{\mu} \nabla_{\nu} \right) f_{R}  = 8 \pi \sum_{i} T^{(i)}_{\mu \nu}  - \sum_{i} f_{T_{i} } \left(T^{(i)}_{\mu \nu} + \Theta^{(i)}_{\mu \nu}\right).
\end{equation}
Here, sum over $\textbf{\textit{i}}$ indicates a sum over the components $\textbf{\textit{m}}$ and $\textbf{\textit{vm}}$. In the conventional $f(R,T)$ theory, one usually assumes the minimal coupling of the form as, $f(R,T)=R+2\mathcal{F}(T)$ \cite{harko2011f}. Similarly, we assume a minimal coupling between matter and geometry, as, $f = R + 2 \mathcal{F} \left(T_{m} ,T_{vm}\right).$ Owing to this choice, $f_{R}=1,$ and the third term on the left hand side of equation (\ref{10}) vanishes. The above equation can then be written as,
\begin{equation}\label{11}
	R_{\mu \nu} - \frac{1}{2}  R \;g_{\mu \nu} = 8 \pi \sum_{i} T^{(i)}_{\mu \nu} + \mathcal{F}\left(T_{m} ,T_{vm}\right) g_{\mu \nu} - \left\{  2 \sum_{i} \mathcal{F}_{T_{i}}  \left(T_{m} ,T_{vm}\right) \left[ T^{(i)}_{\mu \nu} + \Theta^{(i)}_{\mu \nu}\right] \right\} .
\end{equation}

We will now consider the cosmic components as non-interacting with each other, and consequently, we choose, $\mathcal{F}\left(T_{m} ,T_{vm}\right)= \lambda T_{vm} + \kappa T_{m}, $ where $\lambda$ and $\kappa$ are constant coupling parameters. The CDM component having trace $T_m$ is modeled as a pressure-less, perfect fluid having energy momentum tensor,
\begin{equation}\label{12}
	T^{(m)}_{\mu \nu} = \rho_m u_{\mu} u_{\nu}, 
\end{equation}
while the vDM component corresponding to trace $T_{vm}$ has an energy momentum tensor of the form,
\begin{equation}\label{13}
	T^{(vm)}_{\mu \nu} = \left(\rho_{vm} + p_{vm}\right) u_{\mu} u_{\nu} + p_{vm} g_{\mu \nu}.
\end{equation}
Here, $p_{vm} = p^0_{vm} + \Pi$, is regarded as the effective pressure of vDM component, having $p^0_{vm}$ as the equilibrium kinetic pressure and $\Pi$ as the bulk viscous pressure. Also, the kinetic pressure term obeys barotropic equation of state $p^0_{vm} = \omega \rho_{vm}$ with a constant $\omega.$ 

For determining $\Theta^{(i)}_{\mu \nu}$, we use (\ref{8}), (\ref{4}) and also consider the ansatz $ L_{(i)} = -p_i$ \cite{harko2011f} for the matter Lagrangian. Here, $p_i$ is the pressure of the $i^{th}$ component. With these results we obtain the effective field equation as,
\begin{equation}\label{14}
	R_{\mu \nu} - \frac{1}{2}  R \;g_{\mu \nu} = 8 \pi \left[ \bar{T}^{(m)}_{\mu \nu} + \bar{T}^{(vm)}_{\mu \nu}\right],
\end{equation}
where,
\begin{equation}\label{15}
	\bar{T}^{(m)}_{\mu \nu} = T^{(m)}_{\mu \nu} + \frac{ \kappa}{8\pi} \left[2 T^{(m)}_{\mu \nu} + T_{m} g_{\mu \nu}\right]
\end{equation}
and 
\begin{equation}\label{16}
	\bar{T}^{(vm)}_{\mu \nu} = T^{(vm)}_{\mu \nu} + \frac{ \lambda}{4\pi} \left[T^{(vm)}_{\mu \nu} + p_{vm} g_{\mu \nu}+ \frac{T_{vm} g_{\mu \nu} }{2}\right]
\end{equation}
are the effective energy momentum tensors of respective fluid components. To gain simplicity in further investigation, we assume that the explicit coupling of the inviscid matter with geometry as close to zero, i.e. $\kappa \approx 0$. This doesn't mean that we are avoiding this component, instead we stipulate that, owing to it's lack of pressure, this component may have only a negligible explicit coupling with the geometry. 

\subsection*{Dynamics of the Universe}

We consider the universe to be spatially flat, homogeneous \& isotropic, and therefore, adopt the FLRW metric to define the line element,
\begin{equation} \label{17}
	ds^{2}= -dt^2+a(t)^2\left(dx^{2}+dy^{2}+dz^{2}\right).
\end{equation}
Here, $a(t)$ represents the scale factor of the universe and 't' is the cosmic time. The Friedmann equations that describe the dynamics of the universe can be obtained by substituting this metric into the field equation (\ref{14}), and using equations (\ref{15}), (\ref{16}), as,
\begin{equation}\label{18}
	3H^{2}=\bar{\rho}_{vm}+ \rho_m=\rho_{vm} +\tilde{\lambda} \left( 3\rho_{vm}-p_{vm}\right)  + \rho_m 
\end{equation}
\begin{equation}\label{19}
	2\dot{H}+3H^{2}=-\bar{p}_{vm}=-\left[p_{vm} +\tilde{\lambda} \left( 3p_{vm}-\rho_{vm} \right)\right].
\end{equation}
Here, $H=\dot{a}/a$ represents the Hubble parameter of the universe, where the over-dot signifies a derivative with respect to the cosmic time $t$. Note that we have re-scaled $\tilde{\lambda}=\lambda c^4/ 8 \pi G$ and have chosen $8 \pi G/c^4=1$. Also the effective energy density of viscous matter turns out ot be $\bar{\rho}_{vm}=\rho_{vm} +\tilde{\lambda} \left( 3\rho_{vm}-p_{vm}\right) $ as evident form equation (\ref{18}) and the corresponding  effective pressure  as seen from (\ref{19}) is, $\bar{p}_{vm}=p_{vm} +\tilde{\lambda} \left( 3p_{vm}-\rho_{vm} \right)$. 

As stated previously, the effective pressure of the vDM component, i.e, $p_{vm},$ takes the form,
\begin{equation}\label{20}
	p_{vm}=p^0_{vm}+\Pi=\omega \rho_{vm} +\Pi.
\end{equation}
Here, $\Pi$ represents the bulk viscous pressure and according to Eckart's theory $\Pi=-3\zeta H$, where $\zeta$ represents the coefficient of bulk viscosity. Accordingly, (\ref{19}) can now be expressed as,
\begin{equation}\label{21}
	2\dot{H}+3H^{2}=- \left[\omega(1+3\tilde{\lambda})-\tilde{\lambda}\right]\rho_{vm} - (1+3\tilde{\lambda})\Pi.
\end{equation}
Note that, we can also re-interpret this equations as,
\begin{equation}\label{22}
	2\dot{H}+3H^{2}=-\left[\bar{p}^{0}_{vm} + \bar{\Pi}\right],
\end{equation}
where,
\begin{equation}\label{23}
	\bar{p}^{0}_{vm}= \left[\omega(1+3\tilde{\lambda})-\tilde{\lambda}\right]\rho_{vm} = \bar{\omega}\rho_{vm}
\end{equation}
is coupled kinetic pressure with a modified equation of state parameter $\bar{\omega}$ and,
\begin{equation}\label{24}
	\bar{\Pi} = (1+3\tilde{\lambda})\Pi,
\end{equation}
is the coupled bulk viscous pressure.

The corresponding continuity equations for the non-interacting cosmic components are as given below. For vDM component it is,
\begin{equation}\label{25}
	\dot{\rho}_{vm}+3H \left( \rho_{vm}+p_{vm} \right)= -\frac{\tilde{\lambda}\left(\dot{\rho}_{vm} - \dot{p}_{vm}\right)}{1+2\tilde{\lambda}},
\end{equation}
or alternatively, 
\begin{equation}\label{26}
	\dot{\bar{\rho}}_{vm}+3H \left( \bar{\rho}_{vm}+\bar{p}_{vm} \right)= 0.
\end{equation}
And for CDM we have,
\begin{equation}\label{27}
	\dot{\rho}_m+3H\rho_m=0.
\end{equation}
For CDM component, using (\ref{27}) and remembering that $H=\dot{a}/a$, immediately implies the solution,
\begin{equation}\label{28}
	\rho_m = \rho^{0}_m a^{-3}
\end{equation}
where, $\rho^{0}_{m}$ is the present value of CDM density. Note that, all the equations mentioned in this section reduce to the conventional equations of the standard model, if one sets $\lambda= \kappa =0$. We will now proceed to obtain the constraints on model parameters subjected to NEC, CEC and SLT requirements.

\section{Constraints on model parameters due to NEC and CEC} \label{section 3}

In this section we develop the general constraints on the model parameters, for a late accelerating universe, subjected to NEC and CEC (critical energy condition). 
\begin{itemize}
	\item NEC, as mentioned before, demands that the magnitude of bulk viscous pressure of the fluid is much less than its equilibrium kinetic pressure, i.e, $|\Pi/p^0_{vm}|\ll1.$
	\item CEC means, the energy density of vDM should be non-negative ($\rho_{vm}\geq0$). 
\end{itemize}
In imposing these constraints we consider, both negative and positive possibilities for viscous pressure. In almost all literatures in Einstein gravity, only the first choice (i.e, $\Pi<0$) is used, primarily because, only a negative $\Pi$ can generate late acceleration while at the same time satisfy the SLT. However, in the present context, the second possibility, $\Pi>0,$ which corresponds to negative viscous coefficient, cannot be ruled out. This is because of the fact that a positive $\Pi$ can have a negative minimal coupling to gravity, causing a negative effective coupled viscous pressure ($\bar{\Pi}$) and still cause the late accelerated expansion. That is, from (\ref{24}), if $\tilde{\lambda}<-1/3$, then $\bar{\Pi}$ can be negative, even if $\Pi>0$. In such a case, the negativity of $\bar{\Pi}$ will be due to negative nature of coupling parameter $\tilde{\lambda}$. However, to accept the viability of having a positive viscous pressure in this modified gravity, one must also investigate the entropy evolution equation associated with vDM component and check whether it satisfies the SLT, which we will do in Sec. \ref{section 4}. 

\subsection{Constraints based on NEC} \label{section 3 a}

To formulate the constraints, let us first rewrite the NEC given in (\ref{1}) in a more convenient form as,
\begin{equation}\label{29}
	-1\ll \frac{\Pi}{p^0_{vm}} \ll 1
\end{equation}
By combining the Friedmann equations (\ref{18}) and (\ref{19}) we can have the acceleration equation as,
\begin{equation}\label{30}
	\frac{\ddot{a}}{a} = -\frac{1}{6}\left( (3+8\tilde{\lambda})(p^0_{vm}+\Pi) +\rho_{vm} +\rho_m \right).
\end{equation}
Now, imposing the condition for accelerated expansion, $\ddot{a}/{a}>0,$ leads to
\begin{equation}\label{31}
	(3+8\tilde{\lambda})(p^0_{vm}+\Pi) +\rho_{vm}  +\rho_m<0,
\end{equation}
with $\rho_{m} >0$ and $\rho_{vm} >0$ always. Notice that the above inequality depends on the sign of $(3+8\tilde{\lambda})$ and based on that, we can have two distinct cases: 

\subsubsection{\textbf{For $(3+8\tilde{\lambda})<0$ (or $\tilde{\lambda}<-3/8$)}}
Dividing both sides of (\ref{31}) by $3+8\tilde{\lambda}$ causes a flip in the inequality since, $\tilde{\lambda}<-3/8$. It can then be re-written as,
\begin{equation}\label{32}
	-\Pi <p^0_{vm} + \frac{\left( \rho_{vm}  + rho_m \right)}{3+8\tilde{\lambda}}.
\end{equation}
Further, a division by $p^0_{vm}$ on both sides results to,
\begin{equation}\label{33}
	-\frac{\Pi}{p^0_{vm}} < 1 + \frac{\left(1+\frac{\rho_{m}}{\rho_{vm}}\right)}{\omega(3+8\tilde{\lambda})}.
\end{equation} 
where $\omega = p^0_{vm}/\rho_{vm} > 0.$ For $\Pi<0$, the left side of this inequality is strictly positive. For NEC to be satisfied, value of the second term on the right side of the above inequality should be less than 0. However, since the left side of inequality is strictly positive, we cannot have a negative number in the right hand side. Hence, the only possible values of $(1+\rho_{m}/\rho_{vm})/(\omega(3+8\tilde{\lambda}))$, where NEC remains be satisfied is between 0 and -1. Then, for the case where $\Pi>0,$ we express the above inequality as, 
\begin{equation}\label{34}
	\frac{\Pi}{p^0_{vm}} > -1 - \frac{\left(1+\frac{\rho_{m}}{\rho_{vm}}\right)}{\omega(3+8\tilde{\lambda})}.
\end{equation}
This is done to ensure positivity in the left side of the inequality. Notice that, according to (\ref{34}), the NEC is not necessarily violated during the accelerated expansion, if the value of second term on the right hand side is greater than -2.

For satisfying NEC throughout the expansion, one also need to constrain the model parameters by demanding a decelerated expansion of the universe. Interestingly, even though the inequalities ((\ref{33}) and (\ref{34})), were formulated by demanding accelerated expansion for the universe, the same inequalities can be used to investigate the validity of NEC during the prior deceleration phase. For a decelerating universe, one requires $\ddot{a}/a<0$ and in this context (\ref{31}) flips direction, which will in turn flip the inequalities (\ref{33}) and (\ref{34}). This then leads to the following two cases.
\begin{equation}\label{35}
	\textbf{For } \Pi<0 :\;\;\;\;\; -\frac{\Pi}{p^0_{vm}} > 1 + \frac{\left(1+\frac{\rho_{m}}{\rho_{vm}}\right)}{\omega(3+8\tilde{\lambda})}.
\end{equation}
\begin{equation}\label{36}
	\textbf{For } \Pi>0 :\;\;\;\;\; \frac{\Pi}{p^0_{vm}} < -1 - \frac{\left(1+\frac{\rho_{m}}{\rho_{vm}}\right)}{\omega(3+8\tilde{\lambda})}.
\end{equation}
Analyzing these expressions, we see that NEC in the case $\Pi<0$ is not necessarily violated, if the second term on right side is constrained to a value less than 0. Similarly, NEC in the case $\Pi>0$ will be satisfied if value of the second term in the right side is greater than -2. However, since left side of (\ref{36}) is strictly positive, the second term on the right hand side must have a value less than -1. Hence, for satisfying NEC during decelerated expansion in the case where $\Pi>0$, we must constrain the value of $(1+\rho_{m}/\rho_{vm})/(\omega(3+8\tilde{\lambda}))$ between -1 and -2.

In short, in order to satisfy NEC associated with vDM throughout the expansion, we must constrain the value of $(1+\rho_{m}/\rho_{vm})/(\omega(3+8\tilde{\lambda}))$ to be; between 0 and -1 for case $\Pi<0$ and; between -2 and -1 for the case $\Pi>0$. 

\subsubsection{\textbf{For $(3+8\tilde{\lambda})>0$ (or $\tilde{\lambda}>-3/8$)}}
Rearrange the inequality (\ref{31}) and dividing both side by $(3+8\tilde{\lambda})$ we obtain,
\begin{equation}\label{37}
	-\Pi > p^0_{vm} + \frac{\left( \rho_{vm}  + \rho_m \right)}{3+8\tilde{\lambda}}.
\end{equation}
Since $(3+8\tilde{\lambda})>0$, the inequality doesn't flip direction.
If both sides of this inequality is now divided by $p^0_{vm}$, we obtain,
\begin{equation}\label{38}
	-\frac{\Pi}{p^0_{vm}} > 1 + \frac{\left(1+\frac{\rho_{m}}{\rho_{vm}}\right)}{\omega(3+8\tilde{\lambda})}.
\end{equation}  
In this case, since both $\omega$ and $3+8\tilde{\lambda}$ are greater than zero, and the second term on right side remains positive definite. This makes the overall term on the right hand side of the above inequality to be always greater than one and hence, the NEC is violated at all times. That is, for $\Pi<0$, the ratio $-\Pi/p^0_{vm}$ will always be greater than 1 and for the case $\Pi>0$, the ratio $\Pi/p^0_{vm}$ will always be less than -1, which is of course unacceptable. Hence, during accelerated expansion of the universe, there is no such scenario where NEC is satisfied for the vDM fluid having $\tilde{\lambda}>-3/8$ and  $\omega>0$.

\subsection{Constraints based on CEC} \label{section 3 b}
Here we obtain the constraints corresponds to CEC, i.e. $\rho_{vm} >0$ (or equivalently $\Omega_{\rho_{vm}}>0$). To obtain the constraints, we use the Friedmann equation obtained by combining (\ref{18}) with (\ref{20}), given as,
\begin{equation}\label{39}
	3H^{2}=[1+\tilde{\lambda}(3-\omega)]\rho_{vm} -\tilde{\lambda}\Pi + \rho_{m}.
\end{equation}
On rearranging the above equation we get,
\begin{equation}\label{40}
	\Omega_{\rho_{vm}}=\frac{\rho_{vm}}{3H^{2}}=\frac{1+\tilde{\lambda}\Omega_{\Pi}-\Omega_{\rho_m}}{1+\tilde{\lambda}(3-\omega)},
\end{equation}
where, $\Omega_{\rho_m}=\rho_m/3H^2$ and $\Omega_{\Pi}=\Pi/3H^{2}$. Then, the CEC requirement $\Omega_{\rho_{vm}}>0$, implies the constraint,
\begin{equation}\label{41}
	\frac{1+\tilde{\lambda}\Omega_{\Pi}-\Omega_{\rho_m}}{1+\tilde{\lambda}(3-\omega)}>0.
\end{equation}
Analyzing the above inequality we learn that, CEC can hold for the following two distinct cases.
\begin{itemize}
	\item For $1+\tilde{\lambda}\Omega_{\Pi}-\Omega_{\rho_m}>0$ and $1+\tilde{\lambda}(3-\omega)>0$, we get the corresponding constraints as,
	\begin{equation}\label{42}
		\Omega_{\Pi}<\frac{\Omega_{\rho_m}-1}{\tilde{\lambda}} \text{ \; and \; } 	(3-\omega)<\frac{-1}{\tilde{\lambda}}
	\end{equation}
	\item For $1+\tilde{\lambda}\Omega_{\Pi}-\Omega_{\rho_m}<0$ and $1+\tilde{\lambda}(3-\omega)<0$, we get the constraints as,
	\begin{equation}\label{43}
		\Omega_{\Pi}>\frac{\Omega_{\rho_m}-1}{\tilde{\lambda}} \text{ \; and \; } 	(3-\omega)>\frac{-1}{\tilde{\lambda}}
	\end{equation}
\end{itemize}
It is interesting to note that, for $\omega \in (0,1]$ and $\tilde{\lambda}<0$, both of the above cases are applicable. Whereas, for same value of $\omega$ but with $\tilde{\lambda}>0$, only the second case is valid because $3-\omega \nless -1/\tilde{\lambda}$, for any $\tilde{\lambda}>0,$ hence we omit it. Since, we are concerned with scenario where NEC is satisfied, which occurs for $\tilde{\lambda}<-3/8$, both these inequalities are applicable.

\section{\textbf{Entropy Production and Second Law of thermodynamics}} \label{section 4}

In this section we will consider the entropy generation and the consequent constraints if any on the parameters. In the context of (3+1) Einstein's gravity, for the validity of SLT, the viscous coefficient must be greater than zero (i.e, $\zeta>0$)\cite{brevik2013universe}. Here, we will show that, such a condition is not mandatory in the present model. The First law of thermodynamics is given by, 
\begin{equation}\label{44}
	dQ=dU+pdV,
\end{equation}
where $dQ=\mathcal{T}dS$ is the heat energy that enters or leaves the system, $dU$ is the total change in the internal energy of the system and $pdV$ is the amount of work done by (or on) the system. Here, $\mathcal{T}$ is the temperature and $dS$ is the change in entropy. For the vDM component, contained in the three-space volume $V=V_0 a^3$, with total internal energy $U=\rho_{vm} V,$ the first law takes the form,
\begin{equation}\label{45}
	\mathcal{T}dS=\left(\rho_{vm}+p^0_{vm}\right)dV + Vd\rho_{vm}.
\end{equation}
Using the definition of particle number density, $n=N/V$ we can rewrite (\ref{45}) as,
\begin{equation}\label{46}
	\frac{\mathcal{T}dS}{N}=\frac{-\left(\rho_{vm}+p^0_{vm}\right)}{n^2}dn + \frac{d\rho_{vm}}{n}.
\end{equation}
This equation can be further modified to a relation for the entropy production rate as,
\begin{equation}\label{47}
	\mathcal{T}n\dot{s}=\frac{-\left(\rho_{vm}+p^0_{vm}\right)}{n}\dot{n} + \dot{\rho}_{vm}.
\end{equation}
Here, $\dot{s}=(1/N)(dS/dt),$ is the entropy generated per particle. We assume that the particle four-current satisfies the condition $n^{\alpha}_{\; ; \alpha}=0$ \cite{Brevik:2005bj, PhysRev.58.919}, because a violation of this leads to particle creation process \cite{PhysRevD.90.044067} which gives non-equilibrium description for the evolution of the fluid. Since we require the fluid to have a near-equilibrium state, we consider the entropy production in the present model as arising due to the combination of two processes; non-adiabaticity in the expansion \cite{PhysRevD.99.064027} (arising from minimal coupling) and the dissipative effect of bulk viscosity. As a result of these, the number density evolves as,
\begin{equation}\label{48}
	\dot{n}+3Hn=0.
\end{equation}
Using (\ref{25}) and (\ref{48}), we can re-write (\ref{47}) as,
\begin{equation}\label{50}
	\mathcal{T}n\dot{s} = -\left\{\frac{-\tilde{\lambda}\left( \dot{p}_{vm} - \dot{\rho}_{vm}\right)}{1+2\tilde{\lambda}} + 3\Pi H\right\}.
\end{equation}
For a fluid to satisfy SLT, it requires $S^{\alpha}_{\; ;\alpha}\geq0$, where $S^{\alpha}$ is the entropy flow vector. In Eckart theory, entropy flow vector is defined as, $S^{\alpha}=s n^{\alpha}$. It is then easy to see that, given (\ref{48}), we can have $S^{\alpha}_{\; ;\alpha} = n \dot{s}$. Hence, for SLT to be satisfied, it then requires $n \dot{s} \geq 0$ and for this, the entire quantity inside the brackets on the right hand side should be negative. This in turn can be satisfied with $\Pi<0$ or $\Pi>0$. \textbf{For $\Pi<0$}, The overall term inside the curly brackets can be negative in two different ways, (a) $\tilde{\lambda}\left( \dot{p}_{vm} - \dot{\rho}_{vm}\right)/(1+2\tilde{\lambda}) >0$ and (b) $\tilde{\lambda}\left( \dot{p}_{vm} - \dot{\rho}_{vm}\right)/(1+2\tilde{\lambda}) <0$ provided that it's norm satisfies, $|\tilde{\lambda}\left( \dot{p}_{vm} -\dot{\rho}_{vm}\right)/(1+2\tilde{\lambda})| < |3\Pi H|.$ \textbf{For $\Pi>0$}, the entropy production is positive if, $|\tilde{\lambda}\left( \dot{p}_{vm} - \dot{\rho}_{vm}\right)/(1+2\tilde{\lambda})| > |3\Pi H|,$ provided  $\tilde{\lambda}\left( \dot{p}_{vm} - \dot{\rho}_{vm}\right)/(1+2\tilde{\lambda})>0.$ 

As we have already discussed (please refer Sec. \ref{section 3 a}), both $\Pi<0$ and $\Pi>0$ are compatible for explaining the late accelerating epoch while satisfying NEC for vDM, and from above equations, it is clear that both these cases are also compatible with SLT. Note that, while extracting values of model parameters during data analysis, we will consider these conditions also.

\section{Evolution of Hubble parameter } \label{section 5}

In this section we obtain the exact evolution of the Hubble parameter by considering a suitable functional form for the coefficient of bulk viscosity. In literatures, coefficient of bulk viscosity is most often assumed to be function of the expansion rate, it's time derivative and/or energy density. But in general, $\zeta$ can be a function of all these variables, i.e., $\zeta \sim  \zeta(H,\dot{H},\rho_{vm})$. Recently \cite{Gomez:2022qcu}, a parameterized form of bulk viscous coefficient of the form $\zeta \sim \zeta_j H^{1-2j}\rho_{vm}^{j}$ was proposed, where $j$ is a suitable number. This general model is shown to offer a class of viscous forms depending on the value of the constant number $j$. Some popular viscous models such as $\zeta \propto H$ and $\zeta \propto \sqrt{\rho_{vm}}$, can also arise as special cases of this form. In the present analysis, we consider the ansatz, $\zeta=\zeta_0 H + \zeta_1 \rho_{vm} H^{-1}, $ where $\zeta_0$ and $\zeta_1$ are constants, the values of which are to be determined by comparing model with observational data. Here, the first term corresponds to the $j=0$ case and the second term corresponds to $j=1.$ In addition to this, we will also consider a special case of the viscous coefficient obtained from our general ansatz, by setting $\zeta_{0}=0.$ 

\subsection*{Case I: with $\zeta=\zeta_{0}H+\zeta_{1}\rho_{vm}/H$}

The total pressure $p_{vm}$ can be determined by using the relation for viscous pressure $\Pi=-3\zeta H,$ in equation (\ref{20}). Having this, the Friedmann equations (\ref{18}) and (\ref{21}), now takes the form, 
\begin{equation}\label{51}
	3H^{2}=\rho_{vm}\left[1+\tilde{\lambda} \left( 3-\omega + 3 \zeta_{1}\right) \right] + 3\tilde{\lambda}\zeta_0 H^2 + \rho_{m}
\end{equation}
\begin{equation}\label{52}
	2\dot{H}+3H^{2}= \left[ \tilde{\lambda}-( 1 + 3\tilde{\lambda}) \left(\omega-3\zeta_{1}\right)\;\right]\rho_{vm} + 3\zeta_{0}( 1 + 3\tilde{\lambda})H^2.
\end{equation}
Combining the above two equations and substituting for vDM density using (\ref{40}), we obtain the first order differential equation in Hubble parameter as,
\begin{equation}\label{53}
	\frac{2\dot{H}}{3}+ \left\{\frac{(1+2\tilde{\lambda})\left[1+\omega-3\zeta_{1}-(1+4\tilde{\lambda})\zeta_{0}\right]}{1+\tilde{\lambda}\left(3-\omega+3\zeta_{1}\right)}\right\} H^{2}= \left[ \frac{(\omega-3\zeta_{1})(1+3\tilde{\lambda})-\tilde{\lambda}}{1+\tilde{\lambda}\left(3-\omega+3\zeta_{1}\right)}\right]\frac{\rho_{m}}{3}
\end{equation}
Now, substituting for $\rho_{m}$ using (\ref{28}) and making a change of variables from $`t'$ to $'a',$ we then get,
\begin{multline}\label{54}
	\frac{2aH}{3}\frac{dH}{da} + \left\{\frac{(1+2\tilde{\lambda})\left[1+\omega-3\zeta_{1}-(1+4\tilde{\lambda})\zeta_{0}\right]}{1+\tilde{\lambda}\left(3-\omega+3\zeta_{1}\right)}\right\} H^{2} =\\ \left[\frac{(\omega-3\zeta_{1})(1+3\tilde{\lambda})-\tilde{\lambda}}{1+\tilde{\lambda}\left(3-\omega+3\zeta_{1}\right)}\right] \frac{\rho^{0}_{m}a^{-3}}{3}
\end{multline}
The solution of this equation gives the evolution of the Hubble parameter as,
\begin{equation}\label{55}
	H=H_{0}\sqrt{\bar{\Omega}^0_{m} a^{-3}+  \left(1-\bar{\Omega}^0_{m}\right)a^{-\beta}}
\end{equation}
Where $\bar{\Omega}^0_{m}$ and $\beta$ are constants given by,
\begin{equation}\label{56}
	\bar{\Omega}^0_{m}=  \frac{\Omega^0_{m}}{1+\eta} 
\end{equation}
\begin{equation}\label{57}
	\beta=\frac{3(2\tilde{\lambda}+1)\left[(\omega+1)-3\zeta_{1}-(1+4\tilde{\lambda})\zeta_{0}\right]}{1+\tilde{\lambda}(3-\omega+3\zeta_{1})}.
\end{equation}
where, $\eta $ and $\Omega^0_{m}$ are defined as,
\begin{equation}\label{58}
	\eta = \frac{(1+2\tilde{\lambda})(1+4\tilde{\lambda})\zeta_{0}}{\tilde{\lambda} - \omega - 3\tilde{\lambda}\omega + 3(1+3\tilde{\lambda})\zeta_{1}}
\end{equation} 
\begin{equation}\label{59}
	\Omega^0_{m}= \frac{\rho^{0}_{m}}{3H^2_0}.
\end{equation}

From (\ref{55}), it is clear that the Hubble parameter depends on the scale factor through two terms, $a^{-3}$ and $a^{-\beta}.$ The first term is associated with evolution of CDM component, and the second term describes the evolution of vDM. For this model to predict a transition from prior deceleration to a late accelerating era, it is imperative that, $\beta<2$. It is to be noted that, for $\beta <0$ the model become phantom like, for $\beta=0$ it becomes $\Lambda$CDM like and for $0<\beta<2$ model shows quintessence behaviors. 

\subsection*{Case Ia: $\zeta=\zeta_{1}\rho_{vm}/H$}
This special case arises if we assume $\zeta_0=0$ in the previous model. The Hubble parameter in this case takes the simplified form,
\begin{equation}\label{60}
	H=H_{0}\sqrt{\Omega^0_{m} a^{-3}+  \left(1-\Omega^0_{m}\right)a^{-\beta}}
\end{equation}
Since $\zeta_0=0,$ the parameter $\eta$ vanishes hence from (\ref{55}) we then have $\Omega^0_{m}=\bar{\Omega}^0_{m}$ and also the parameter $\beta$ becomes,
\begin{equation}\label{61}
	\beta=\frac{3(2\tilde{\lambda}+1)\left[(\omega+1)-3\zeta_{1}\right]}{1+\tilde{\lambda}(3-\omega+3\zeta_{1})}.
\end{equation}
Even though this model also shows a late accelerated expansion of the universe for ($\beta<2$), the dynamics of this model is notably different from the previous one, as will be evident (later on) from the evolution of cosmological parameters. But unlike the previous model, the present one will not reduces to $\Lambda$CDM if $\beta=0$ (refer equation (\ref{61})). The case $\beta=0,$ implies two possibilities, i.e, $\tilde{\lambda}=-1/2$ or $\zeta_{1}=(1+\omega)/3$. First case is immediately ruled out since, it causes (\ref{50}) to diverge. While in the second case, we learn that, by setting $\zeta_{1}=(1+\omega)/3$, the Hubble parameter given in equation (\ref{60}) becomes independent of coupling parameter $\tilde{\lambda}$, because of this, the value of $\tilde{\lambda}$ cannot be determined by comparing model with data. However, the conservation equation (\ref{25}) is seen to depend on value of $\tilde{\lambda}$. This poses a serious issue as value of $\tilde{\lambda}$ is indeterminate in this case. Hence, we can omit $\Lambda$CDM like behavior of this case from further analysis.

\subsection*{Case Ib: $\Lambda$CDM behavior as a particular case}

\begin{table}
	\centering
	\begin{tabular}{|c|c|c|c|}\hline
		Model& Case I &Case Ia& Case Ib\\
		Parameters& & &\\ \hline
		$H_0$&$70.15^{+1.51}_{-1.50}$&$70.16^{+1.32}_{-1.30}$&$70.35^{+1.15}_{-1.17}$ \\
		$\tilde{\lambda}$&$-0.498^{+0.003}_{-0.002}$&$-0.497^{+0.002}_{-0.002}$&$0.88$ \\
		$\alpha$&$0.19^{+0.15}_{-0.16}$&$0.18^{+0.16}_{-0.15}$&$-$ \\ 
		$\Delta$&$0.53^{+0.33}_{-0.30}$&$0.49^{+0.34}_{-0.36}$&$-$ \\
		$\Omega^{0}_{m}$&$0.25^{+0.01}_{-0.01}$&$0.26^{+0.01}_{-0.01}$& $0.27^{+0.17}_{-0.15}$\\ 
		$\zeta_{0}$&$0.19$&$-$&$-0.003^{+0.02}_{-0.03}$ \\
		$\zeta_{1}$&$-0.22$&$-0.23$&$0.50^{+0.32}_{-0.30}$ \\ 
		$\omega$&$0.87^{+0.12}_{-0.13}$&$0.87^{+0.11}_{-0.12}$&$0.48^{+0.34}_{-0.30}$ \\ 
		M&$19.35^{+0.02}_{-0.02}$&$19.33^{+0.02}_{-0.02}$&$19.34^{+0.01}_{-0.01}$ \\
		$\chi^{2}_{min}$&$1069.12$&$1068.94$&$1075.94$ \\
		$\chi^{2}_{dof}$&$0.979$&$0.978$&$0.983$ \\ \hline
	\end{tabular}
\caption{Best estimated values of model parameters for each case by applying the respective model constraints and comparing models to OHD+SNe Ia data sets.}\label{tab1}
\end{table} 

The Hubble parameter for the case I, as given in (\ref{55}), shows an exact $\Lambda$CDM behavior when 
$\beta=0$. According to (\ref{57}), this can be realized in two different ways. First case is with $\tilde{\lambda}=-1/2$ and the second is with $\tilde{\lambda}=\epsilon=\left\{\left[(\omega+1)-3\zeta_{1}\right]/\zeta_{0}-1\right\}/4$. For $\tilde{\lambda}=-1/2$, as explained in Case Ia, the entropy production as given by equation (\ref{50}) will diverge and becomes indeterminate, hence, we neglect this scenario from further analysis. However, this is not the case for, $\tilde{\lambda}=\epsilon.$ Here, we see that, considering $\tilde{\lambda}=\epsilon$ in (\ref{58}) and (\ref{56}), leads to a new definition for $\bar{\Omega}^0_{m}$ as,
\begin{equation} \label{62}
	\bar{\Omega}^0_{m}= \left[\frac{4\left(1-\omega+3\zeta_{1}\right)}{3-\omega+3\zeta_{1}+\zeta_{0}} -1\right]\Omega^0_{m}.
\end{equation}
Accordingly, the evolution of Hubble parameter in (\ref{57}) reduces to $\Lambda$CDM like behavior
\begin{equation}\label{63}
	H=H_{0}\sqrt{\bar{\Omega}^0_{m} a^{-3}+  \left(1-\bar{\Omega}^0_{m}\right)},
\end{equation}
but with a modified dark matter density. This suggest that, the vDM component having a minimal coupling to gravity which satisfies the condition $\tilde{\lambda}=\epsilon$, is capable of showing cosmological constant like behavior in the late phase of the universe.

\section{Evolution of Cosmological observables using Best estimated value of model parameters} \label{section 6}

In the previous section, we obtained the solutions for the Hubble parameter for different cases. Now we will analyze its evolution subjected to NEC, CEC and SLT by applying the constraints that were developed in Sec. \ref{section 3} and Sec. \ref{section 4}.  For this purpose, we first extracted the model parameters, i.e., $\tilde{\lambda}, \omega, \zeta_0, \zeta_1, ....$ etc subjected to the conditions given equations (\ref{33}), (\ref{34}) (corresponding the NEC), (\ref{41}) (which corresponds to CEC) and (\ref{50}) (corresponding to SLT). We use type Ia supernovae data \cite{Scolnic_2018} and the observational Hubble data \cite{Geng:2018pxk} for the computation. To ensure the continuance of smooth discussion, we have given details of the data analysis in appendix \ref{appen}. The best estimated value of model parameters hence obtained are provided in Table \ref{tab1}. 

\subsection{\textbf{Case I}} \label{c1}

\begin{figure}
	\centering
	\includegraphics[width=0.6\columnwidth]{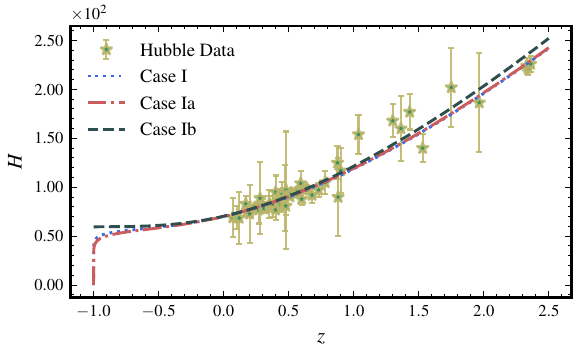}
	\caption{ Graph comparing the exact evolution of Hubble parameter obtained by using best estimated value of model parameters from combined OHD+SNe Ia data sets.}
	\label{OHD}
\end{figure}
\begin{figure}
	\centering
	\includegraphics[width=0.6\columnwidth]{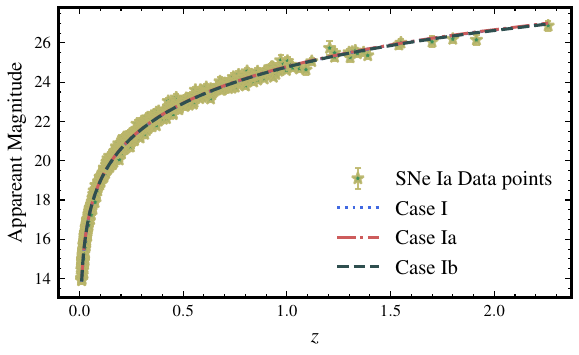}
	\caption{ Graph comparing the value of apparent magnitude with values obtained from it's theoretical model using best estimated value of model parameters.}
	\label{SNeIa}
\end{figure}
\begin{figure}
	\centering
	\includegraphics[width=0.7\columnwidth]{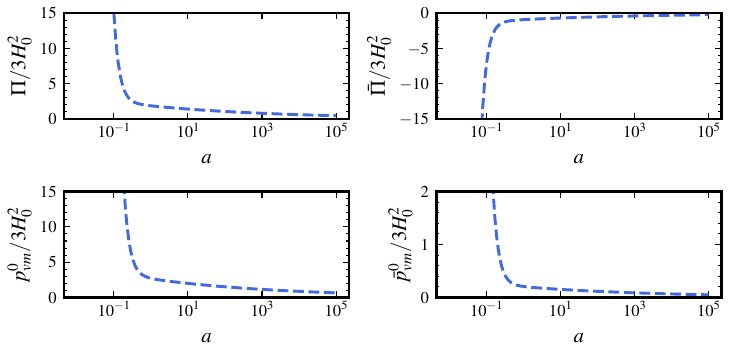}
	\caption{ Evolution of different pressure terms associated with Case I, under best estimated value of model parameters. $p^{0}_{vm}$ and $\Pi$ are the uncoupled kinetic and bulk viscous pressure, while $\bar{p}^0_{vm}$ and $\bar{\Pi}$ represent their respective coupled versions. Note that each pressure is scaled by a factor of $3H^2_{0}$.}
	\label{prssr1}
\end{figure}
\begin{figure}
	\centering
	\includegraphics[width=0.7\columnwidth]{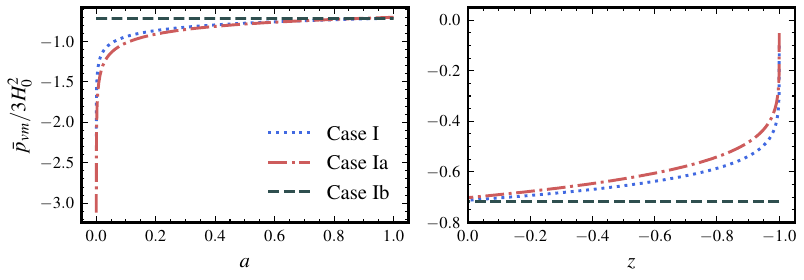}
	\caption{ Evolution of effective pressure $\bar{p}_{vm}$ with scale factor in different models corresponding to the best estimated value of model parameters.}
	\label{peff}
\end{figure}
Fig. \ref{OHD} shows the evolution of the Hubble parameter for all the three cases while, Fig. \ref{SNeIa}, compares the model prediction of the apparent magnitude of the SNe Ia at various redshifts with the corresponding observed values. From the two figures, we notice that, the evolution of $H$ in Case I and Case Ia are not much different from $\Lambda$CDM like behavior shown in Case Ib, so that all of them predicts late acceleration. However, both Case I and Case Ia predict far future quintessence era for the universe, whereas Case Ib depicts an end de-Sitter epoch. Also, since the best estimated value of the viscous coefficient $\zeta_1$ is negative for both Case I and Ia  (refer table (\ref{tab1})), the viscous pressure in both cases are obtained to be positive. One may now ask the obvious question that, "If both viscous pressure and kinetic pressure are positive, what produces the negative pressure that is necessary for the recent cosmic expansion ?".

To answer this, we analyze (\ref{21}), (\ref{22}), (\ref{23}) and (\ref{24}). From which we find that, even though the bulk viscous pressure ($\Pi$) is positive, the coupling of vDM with gravity makes the coefficient $(1+3\tilde \lambda),$ as given in equation (\ref{24}) negative, which results in a negatively coupled bulk viscous pressure $\bar{\Pi}.$ Please see Fig. \ref{prssr1}. It is this negative pressure, that causes the late accelerated expansion of the universe. Despite of having a negative coupling parameter, the coupled kinetic pressure of the viscous fluid, $\bar{p}^0_{vm}$ as in (\ref{23}) is still positive, due to the positiveness of the coupled equation of state, $\bar \omega.$ Also, it is easy to see that for parameter values in the domain $\tilde{\lambda} \in (-1/2,-3/8)$ and $\omega \in (0,1)$, coupled kinetic pressure is always greater than zero and only the coupled viscous pressure can attain a negative value ($\bar{\Pi}<0$).
\begin{figure}
	\centering
	\includegraphics[width=0.6\columnwidth]{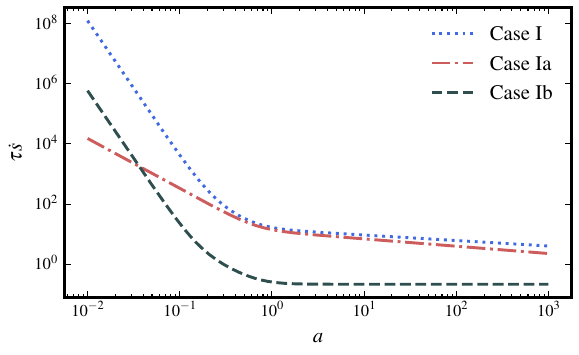}
	\caption{ Evolution of (\ref{50}) plotted against scale factor ($a$) for best estimated value of model parameters. Here, $\tau$ has the definition $\tau = Tn/3H^3_0$.}
	\label{slt}
\end{figure}
Therefore, with a negative coupling constant $\tilde{\lambda}$, the effective viscous pressure of vDM can generate the adequate negative pressure, for causing late accelerated expansion of the universe. In Fig. \ref{peff}, we confirm that the effective pressure of coupled viscous fluid is indeed negative and diverging as $a \to 0$ and later in the far future stage of evolution, (i.e, as $z \to -1$), it asymptotically approaches zero from the negative side. 

To check whether, the case $\Pi>0$ will satisfy the SLT or not, we studied the evolution of entropy production, as given in relation (\ref{50}), with scale factor, for best estimated value of model parameters. The corresponding graph is presented as Fig. \ref{slt}. Analyzing the same, we learn that the entropy rate $\dot{s}$, is indeed positive, and is decreasing asymptotically to zero as $a \to \infty$. At the outset it is evident from this figure that, the entropy is at the increase, even with $\Pi>0.$
\begin{figure}
	\centering
	\includegraphics[width=0.6\columnwidth]{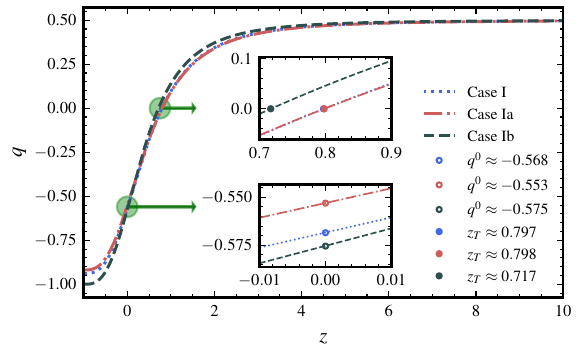}
	\caption{ Evolution of deceleration parameter with redshift for best estimated value of model parameters.}
	\label{dec}
\end{figure}

The nature of evolution of the deceleration parameter ($q$) can be studied using the standard relation,
\begin{equation} \label{64}
	q=-\frac{\ddot{a}}{aH^{2}} = -\left(1+\frac{\dot{H}}{H^2}\right). 
\end{equation}
Fig. \ref{dec} shows it's evolutionary behavior with change in redshift. Accordingly, the transition to late accelerated epoch has occurred at a redshift, around $z_{T}=0.797$ and the present value of deceleration parameter is $q_{0}\approx-0.568$. These values are very close to estimates of the current concordance model. However, $q$ is not approaching the value $-1$ as in the standard $\Lambda$CDM model, but to a value greater than -1. Hence, we conclude that the present case leads to an end quintessence behavior rather than a de-Sitter one.
\begin{figure}
	\centering
	\includegraphics[width=0.6\columnwidth]{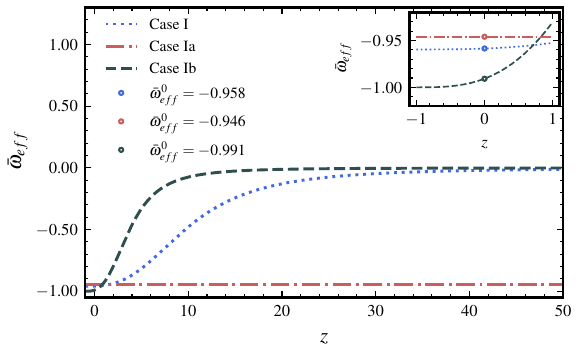}
	\caption{Evolution of effective equation of state ($\bar{\omega}_{eff}$) of minimally coupled vDM component with redshift in each model, for best estimated value of model parameters.}
	\label{wef}
\end{figure}
\begin{figure}
	\centering
	\includegraphics[width=0.6\columnwidth]{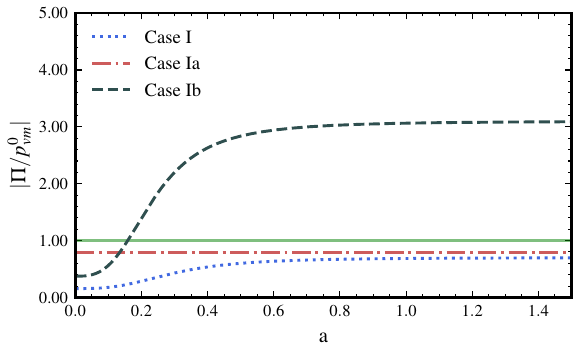}
	\caption{ Variation in NEC with scale factor, graphed using the best estimated value of model parameters. The green line represents the boundary where, $\Pi/p^0_{vm}=1$.}
	\label{NEC}
\end{figure}
\begin{figure}
	\centering
	\includegraphics[width=0.6\columnwidth]{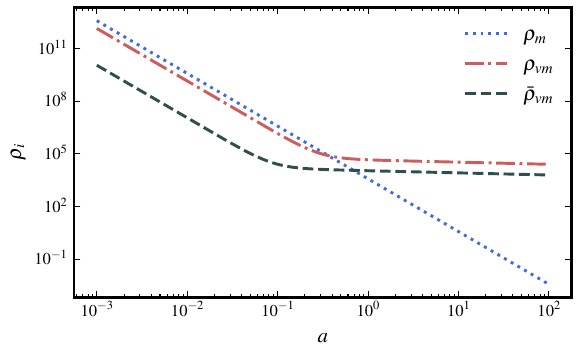}
	\caption{Evolution of energy density associated with each component against change in scale factor for case I. Note that each axis has been scaled logarithmically.}
	\label{density1}
\end{figure}
In Fig. \ref{wef}, we have shown the evolution of the effective equation of state of coupled vDM component which can be obtained from $\bar{p}_{vm}$ and $\bar{\rho}_{vm}$ as,
\begin{equation}\label{65}
	\bar{\omega}_{eff}=\frac{\bar{p}_{vm}}{\bar{\rho}_{vm}}=\frac{\bar{p}^0_{vm}+\bar{\Pi}}{\bar{\rho}_{vm}}.
\end{equation}
It shows that, the effective equation of state of coupled vDM component, evolves from $\bar{\omega}_{eff} \to 0$ as $a \to 0$ to a value greater than -1, as $z \to -1.$ It's present values is found to be around $\bar{\omega}_{eff} \approx -0.958$ which is very close to -1. However, since the effective equation of state of vDM saturates to a value greater than -1, we can again confirm that, coupled vDM is mimicking a quintessence dark energy component.

At this juncture, it is to be noted that the the inherent or barotropic equation of state $\omega$ (refer equation (\ref{23})) of the vDM, corresponding to its kinetic pressure, would have value around $\omega \approx 0.87$ (close to stiff matter like fluid). But due to its minimal coupling with geometry (through the parameter $\tilde{\lambda}$), it attains a 'coupled equation of state' (\ref{23}), which has a value around $\bar{\omega} \approx 0.067$. This shows that, the vDM fluid having an almost stiff matter like equation of state, behaves like a viscous warm dark matter like fluid owing to its negative minimal coupling to gravity. 

We obtained the age of the universe predicted by this model, by integrating the  Hubble parameter (i.e. $dt/da={[aH(a)]}^{-1}$) of the universe as,
\begin{equation}\label{66}
	t_{0}-t_{b}= \int_{0}^{a_0}\frac{da}{aH(a)}. 
\end{equation}
Here, $t_{0}$ represents the present time (corresponds to $a=a_0=1$) and $t_{b}$ is the time at which the big bang (corresponds to $a=0$) occurred. Hence, the interval $t_{0}-t_{b}$ represents the current age of the universe since the big bang. Using the extracted model parameters, the calculated age is approximately around $13.91$ Gyrs, which is very close to $\Lambda$CDM model prediction, but slightly greater. Finally, by analyzing Fig. \ref{NEC} and Fig. \ref{density1}, we confirm that this model is well behaving under NEC and CEC conditions.

\subsection{\textbf{Case Ia}}

\begin{figure}
	\centering
	\includegraphics[width=0.7\columnwidth]{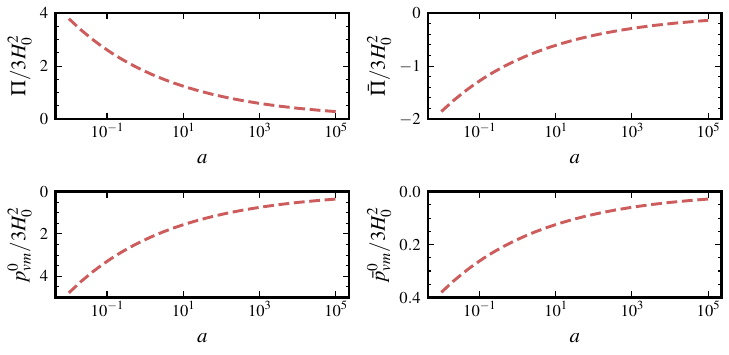}
	\caption{ Evolution of different pressure terms associated with Case Ia, under best estimated value of model parameters.}
	\label{prssr2}
\end{figure}

The evolution of the Hubble parameter in this case, included in Fig.\ref{OHD}, implies a transition into the accelerated expansion at the late stage of the evolution. Behavior of this model is similar to that of case I and is not too much different from that of the $\Lambda$CDM. This model also predicts the magnitudes of the distant supernovae as is evident from Fig. \ref{SNeIa}. However, a distinguishing feature of this case is that, the ratio of the bulk viscous pressure to the kinetic pressure is a constant, which satisfies the relation, $\big|\Pi\big|/p^0_{vm} = 3\zeta_1/\omega$.

From best estimated values of these parameters (and from Fig. \ref{NEC}), it is evident that the above ratio is less than one, which confirms the validity of NEC. This constancy in the ratio of $\Pi$ and $p^0_{vm}$ also implies a similar evolutionary natures for $\Pi$ and $p^0_{vm}$, which can be seen from Fig. \ref{prssr2}. The figure also shows that $\Pi$, $p^0_{vm}$ and $\bar{p}^0_{vm}$ are always positive, while the coupled viscous pressure, which is responsible for the late accelerated expansion of the universe, remains negative. Evolution of the effective pressure, which comprises of the coupled kinetic and bulk viscous pressures of vDM approaches zero from the negative side as $a \to \infty$, as shown in  Fig. \ref{peff}.

Another feature which distinguishes this model from the previous case, is the constancy of the effective equation of state, with its value is around $\bar{\omega}_{eff}= -0.946$. This shows that, the minimally coupled vDM component effectively mimics a quintessence dark energy component with constant equation of state. However, it is to be noted that the barotropic equation of state $\omega,$ of vDM implies a stiff matter like behavior (refer Table (\ref{tab1})), which on coupling with spacetime acquire the character of quintessence fluid of constant negative equation of state.  

The evolution of the deceleration parameter ($q$) given in Fig. \ref{dec}, reveals that the effect of the negative equation of state of vDM will dominate only at the later stage, at which the universe enter an accelerating epoch. The transition to late accelerated epoch is found to have occurred at a redshift of $z_{T}=0.798$ and the present value of $q$ is around $q_{0}\approx-0.553.$ 
\begin{figure}
	\centering
	\includegraphics[width=0.6\columnwidth]{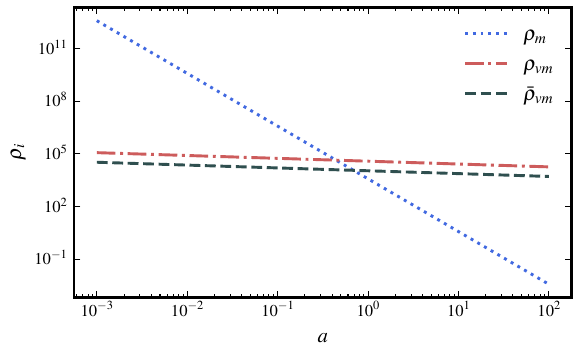}
	\caption{Evolution of different energy density terms with scale factor for case Ia. Notice that, the two lines are not perfectly horizontal, but have slight inclination to them. This suggests that the vDM density is not a constant but is indeed decreasing over time.}
	\label{density2}
\end{figure}

Following this, we estimated the age of the universe predicted by this model, using (\ref{60}) and (\ref{66}), to be around $13.87$ Gyrs which is, very close to the age predicted by $\Lambda$CDM model. Finally, we have analyzed the evolution of density of vDM by plotting Fig. \ref{density2}. It is clear that, the density of vDM is always greater than zero, which confirm the validity of CEC.

\subsection{\textbf{Case Ib}}

In a previous section, we have shown that the model in this special case is dynamically equivalent to the standard $\Lambda$CDM model. Now, let us analyze the behavior of cosmological parameters using the best estimated value of model parameters given in Table \ref{tab1}. Compared to the previous cases (Case I $\&$ Ia), we found that, when NEC constraints are imposed on the system while simultaneously requiring the condition $\tilde{\lambda}=\epsilon$, the density of vDM, i.e, $\rho_{vm}$, becomes negative at certain epochs of the evolution of the universe. It then runs out that, if one lifts the NEC constraints while evaluating the model parameters, this negativity in the energy density disappears and at the same time the model retains $\Lambda$CDM nature. Hence, we investigate this special case without caring the NEC. We point out that, our motivation for considering this case, is just to find, under what conditions the mixed matter component model shows $\Lambda$CDM like behavior.

\begin{figure}
	\centering
	\includegraphics[width=0.6\columnwidth]{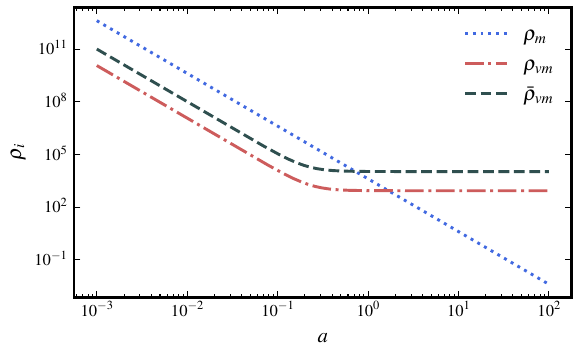}
	\caption{Evolution of energy density associated with each component against change in scale factor for case Ib.}
	\label{density3}
\end{figure}

Fig. \ref{OHD} and Fig. \ref{SNeIa}, shows that this model exhibit good fit to both OHD and SNe Ia data, and asymptotically evolves towards a de Sitter epoch. The transition into the recent acceleration is found to have occurred at a redshift of $z_T = 0.717$ (see Fig. \ref{dec}), while the present value of deceleration parameter is $q_0\approx-0.575$. Fig. \ref{wef} reveals the evolution of the effective equation of state of coupled vDM component, which suggest a decay from $\bar{\omega}_{eff} \to 0$ at $a \to 0$ to $\bar{\omega}_{eff}\to -1$ as $a \to \infty$. Here, we notice that the effective equation of state $\bar{\omega}_{eff}$ approaches zero, as $a \to 0$, much faster than in Case I. From Fig. \ref{NEC}, we see that, even though NEC is not satisfied for the model, it's value saturates at around $|\Pi/p^0_{vm}|\approx 3.1$ in the far future evolution. In addition, from Fig. \ref{density3} and Fig. \ref{slt} we find that model is well behaving under both CEC and SLT conditions.
\begin{figure}
	\centering
	\includegraphics[width=0.7\columnwidth]{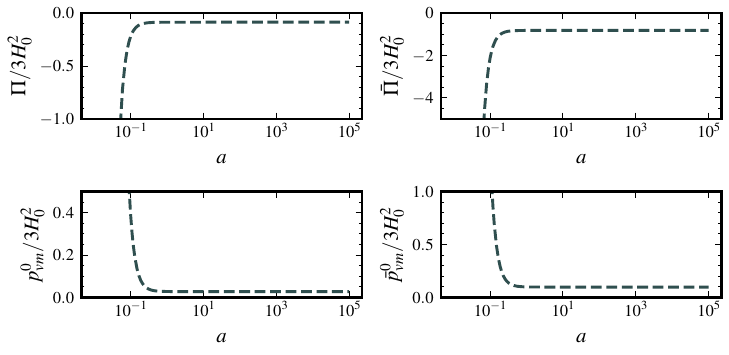}
	\caption{ Evolution of different pressure terms associated with Case Ib, under best estimated value of model parameters.}
	\label{prssr3}
\end{figure}
In Fig. \ref{peff}, it is shown that the effective pressure $\bar{p}_{vm},$ of the coupled fluid remains as a constant throughout the evolution. By analyzing (\ref{65}), we notice that, this is due to the complementary behaviors of the effective equation of state and the density of the coupled dark matter fluid. However, it should be noted that, even if the effective pressure of vDM remains as a constant, the kinetic pressure $\bar{p}^0_{vm}$ and the viscous pressure $\bar{\Pi}$ of the coupled vDM fluid varies as shown in Fig. \ref{prssr3}. In the asymptotic far future, both the effective pressure ($\bar{p}_{vm}$) and effective density ($\bar{\rho}_{vm}$) becomes constant and hence, the net effective equation of state of coupled vDM fluid becomes -1. The age predicted by this model is determined using (\ref{63}) and (\ref{66}) and is found to be 13.61 Gyrs, which is significantly close to the value predicted by the standard model.

\section{Results and Discussion}

In explaining the recent accelerated expansion of the universe using dissipative effects in the matter sector, in the context of Einstein's gravity, it was shown that the presence of cosmological constant is essential to satisfy NEC during the evolution of the viscous fluid, that too for some particular choice of bulk viscous coefficient \cite{Cruz:2022zxe,sym14091866,PhysRevD.105.024047}. However, in such models where both dark energy and viscous matter are considered, the late accelerated expansion is driven mainly by the dark energy component, while the contribution by viscous matter remains significantly small. This means, in Einstein's gravity, it is not possible to have late accelerated expansion of the universe driven by bulk viscous matter whilst maintaining it's near equilibrium state. Hence it is necessary to search, the possibility of generating late acceleration using viscous effects in the matter sector by satisfying the NEC, without the needing dark energy, in a modified gravity context. In the present work, we have explored such a possibility in the context $f(R,T)=R+2\lambda T_{vm}$ gravity. We showed that, by considering mixed dark matter components (non-interacting mixture of viscous DM and inviscid CDM) in the context of $R+2\lambda T_{vm}$ gravity and imposing suitable constraints on the model parameters, it is possible to attain a viscous-driven late accelerated expansion of the universe while simultaneously satisfying the NEC for the dissipative matter, even in the absence of a cosmological constant.

First, we formulated the general constraints on model parameters by imposing the NEC, CEC and SLT requirements on vDM component. The constraints developed are general since, they were derived without assuming any phenomenological form for bulk viscous coefficient. An in depth analysis of the constraints showed that, in this modified gravity regime, the possibilities of simultaneously satisfying NEC, CEC and SLT throughout the expansion exists, only if the coupling parameter has a value in the domain $\tilde{\lambda}\in (-1/2,-3/8)$. Interestingly,  we also noticed that, this was possible for both negative and positive cases of viscous pressures (i.e, with both $\Pi>0$ and $\Pi<0$). One of the intriguing result obtained in this context, which is contrary to the result in Einstein gravity, was the possibility of having a positive viscous pressure for the vDM component while still having a late accelerated expansion of the universe. This result was obtained as a direct consequence of considering a vDM component with a negative coupling to geometry (i.e, with $\tilde{\lambda} \in (-1/2,-3/8)$). By analyzing (\ref{22}), we learned that, for having a late accelerated expansion of the universe generated by bulk viscosity, one must have the coupled viscous pressure $\bar{\Pi}<0$. And from (\ref{23}), we saw that, this can occur for two distinct cases, either with $\Pi<0$ and $\tilde{\lambda}>-1/3$ or with $\Pi>0$ and $\tilde{\lambda}<-1/3$. Even though both cases can equally explain late accelerated expansion of the universe, only the latter case, i.e, $\Pi>0$ with $\tilde{\lambda}<-1/3$, satisfies the NEC requirement with vDM component (refer Sec. \ref{section 3 a}). Then, to confirm the viability of having such a positive viscous pressure in this modified gravity, we investigated the entropy evolution associated with vDM component and showed that, in the context of this modified gravity, it is possible to satisfy SLT associated with vDM component even with $\Pi>0$.

A complete analysis of the model was then carried out by suitably choosing the viscous coefficient, by keeping NEC, CEC and SLT requirements satisfied throughout the expansion. As the primary choice, we have considered the form of viscous coefficient as, $\zeta = \zeta_{1}\rho_{vm}/H + \zeta_{0}H$, which we investigated as Case I, and then considered $\zeta = \zeta_{1}\rho_{vm}/H$ as it's special case, which we labeled as Case Ia. In both these cases, the Hubble parameter model predicted a universe which starts off from an initial big bang singularity, which then undergoes a decelerated expansion followed by a late accelerated epoch which ends with a quintessence/de-Sitter/phantom era in it's far future evolution. However, we found that, a far-future de-Sitter epoch for Case Ia is not favorable as the value of parameter $\tilde{\lambda}$ in this scenario becomes indeterminate from the data analysis. 
For academic curiosity, we have separately analyzed the special case where the Case I shows $\Lambda$CDM like behavior as Case Ib. We then compared each of these models with OHD+SNe Ia data, by applying the model constraints, and obtained the best estimated value of model parameters.

Comparison of the models with observational data showed that, in addition to having a good fit to observational data, Case I and Case Ia satisfies all three necessary conditions (NEC, CEC and SLT) throughout the evolution. Both of these models depicted a late accelerating universe which approaches a quintessence era in their far future evolution. From detailed analysis, we learned that, the recent accelerated expansion in these models is driven by the coupled vDM component and the past deceleration is obtained as a result of CDM density dominating over vDM density. We also learned that, the viscous pressure ($\Pi$) in these cases is positive and hence cannot directly drive the accelerated expansion of the universe. However, owing to a negative minimal coupling with geometry, the effective coupled bulk viscous pressure ($\bar{\Pi}$) becomes negative and hence causes the late-accelerated expansion. Affirming the same, we also saw that, in the absence of viscous pressure ($\Pi$), it is not possible to achieve accelerated expansion in these models with any $\tilde{\lambda} \in (-1/2,-3/8)$ for any $\omega \in (0,1]$. 

\begin{table}
	\centering
	\begin{tabular}{|c|c|c|c|c|} \hline
		Observables&$\Lambda$CDM&Case I&Case Ia & Case Ib \\ \hline
		Age (in Gyrs)&$13.79\text{\cite{refId0}}$&$13.91$&$13.87$&$13.61$ \\
		$q^0$&$-0.55\text{\cite{aghanim2020planck}}$&$-0.568$&$-0.553$&$-0.575$ \\
		$z_{T}$&$0.683\text{\cite{Jesus_2020}}$&$0.797$&$0.798$ & $0.717$\\
		$\bar{\omega}^0_{vm}$&-&$-0.958$& $-0.946$& $-0.991$\\
		$\omega_{DE}$&$-1$&- &-& - \\ \hline
	\end{tabular}
	\caption{Comparison between cosmological observables obtained for each case with values predicted by $\Lambda$CDM model.}\label{tab2}
\end{table}

From analysis of Case Ib, we learned that, the model was successful in predicting the end de Sitter phase for the universe, however, the NEC associated with vDM component had to be violated. Nevertheless, this model obeys CEC requirement and also returns a good fit to data and also predicts values of cosmological parameters close to their standard values. Contrary to previous cases (i.e., Case I and Ia), we found that the viscous pressure in this model is negative and by studying the entropy evolution of vDM component, we showed that this model is also in agreement with SLT. To show a quick comparison between models, we have tabulated the values obtained for cosmological observables corresponding to each case and the standard $\Lambda$CDM values in Table \ref{tab2}.

To sum up, we found that NEC associated with vDM component with viscous coefficient $\zeta = \zeta_{1}\rho_{vm}/H + \zeta_{0}H$ or $\zeta = \zeta_{1}\rho_{vm}/H$, could be maintained throughout the evolution in $R+2\lambda T_{vm}$ gravity even in the absence of cosmological constant or dark energy. The simultaneous NEC and CEC requirement, lead to the scenario where bulk viscous pressure of the vDM component becomes positive, but owing to a negative minimal coupling with geometry, the effective pressure of the vDM fluid becomes negative which then leads to the recent accelerated expansion of the universe. However, we saw that, in this modified gravity, having a positive bulk viscous pressure needn't necessarily violate the SLT, hence allowing the possibility of a negative viscous coefficient. Finally, the models studied based on these constraints show quintessence behavior in their far future evolution and in addition to being a good fit to data, the predicted values of cosmological variables are significantly close to the standard values determined from the current concordance model.
	
\appendix

\section{Data Analysis} \label{appen}

For estimating best fit value of model parameters we compare analytical models with observational data. For this we have chosen, Type Ia Supernovae data (SNe Ia) \cite{Scolnic_2018} which contains a total of 1048 data points within a redshift range of $0.01\leq z \leq2.26$ and the Observational Hubble data (OHD) \cite{Geng:2018pxk} which contains 51 data points within redshift range of $0.07\leq z \leq 2.36$. Comparison of each model with combined data set is done using standard $\chi^{2}$ analysis employed through Markov chain Monte Carlo (MCMC) estimation technique by utilizing emcee python package \cite{Foreman-Mackey_2013} in lmfit python library \cite{newville_matthew_2014_11813}. Since the constraints on free parameters are expressions in the form of inequalities, they are implemented using expression bound techniques available in lmfit library.

For our analysis using the OHD data, we compare the values of theoretical Hubble parameter $H_{t}$, obtained for different redshifts, with those in the observational Hubble data $H_{o}$ which are also measured at different redshifts. The required $\chi^{2}$ function which is to be minimized is then given by,
\begin{equation}\label{a1}
	\chi^{2}_{OHD}((a,b,. ,H_{0})) = \sum_{k=1}^{n}\frac{\left[H_{t}(a,b,. ,H_{0})-H_{o}\right]^{2}}{\sigma_{k}^{2}}
\end{equation}

Where, $a,b,..,H_{0}$ represents the model parameters whose best estimates are to be found, $n$ is the total number of data points available for the analysis and $\sigma_{k}^{2}$ is the variance in the measured value of  $k^{th}$ data. For comparing the models with type Ia supernovae data (SNe Ia), we use the theoretical expression for the distance modulus $\mu_{t}$ of $k^{th}$ supernovae at red shift $z_{k}$, given by, 
\begin{align}\label{a2}
	\mu_{t}(z_{k},a,b,.,H_{0})&=m-M\\
	&=5\log_{10}\left[\frac{d_{L}(z_{k},a,b,.,H_{0})}{\text{Mpc}}\right]+25. \notag
\end{align}
Here, $m$ and $M$ are apparent and absolute magnitudes of supernovae and $d_{L}$ is the luminosity distance defined for a flat universe. Through out the analysis we treat $M$ as a nuisance parameter. The relation for luminosity distance is then given by, 
\begin{equation}\label{a3}
	d_{L}(z,a,b,.,H_{0})=c(1+z)\int_{0}^{z}\frac{dz'}{H(z',a,b,.,H_{0})}.
\end{equation}

Hence, the required $\chi^{2}$ function is,
\begin{equation}\label{a4}
	\chi^{2}_{SNe}((a,b,.,H_{0}))  = \sum_{k=1}^{n}\frac{\left[\mu_{t}(a,b,.,H_{0})-\mu_{o}\right]^{2}}{\sigma_{k}^{2}}.
\end{equation}
For the combined data analysis using both OHD and SNe Ia data sets, the $\chi^{2}$ function to be minimized is given by,
\begin{equation}\label{a5}
	\chi^{2}_{total} = \chi^{2}_{OHD} +\chi^{2}_{SNe} .
\end{equation}

Using these equations, we perform $\chi^{2}$ minimization for the each model and estimate the best fit values of all model parameters. The viability of a given model based on the observational data is determined by finding the chi square per degrees of freedom, which is defined as,
\begin{equation}\label{a6}
	\chi^{2}_{dof}=\frac{\chi^{2}_{min}}{n-n_{p}}.
\end{equation}
Here, $n$ is the number of available data points and $n_{p}$ is the number of model parameters. The model is considered as, good fit to data if $\chi^{2}_{dof} \approx 1$, over fits the data if $\chi^{2}_{dof}\ll1$ and bad fit to data if $\chi^{2}_{dof}\gg1$. 

For restricting values of model parameters in each model, we will consider constraints developed in Sec. \ref{section 3} with suitable modifications and we will also mention the domain of values that were chosen for bounding each parameter. From our detailed analysis using different sets of priors, we have reported only those cases that satisfy NEC, CEC and SLT simultaneously, except for Case Ib, where we have neglected NEC and varied the parameters freely.

\paragraph*{\textbf{Case I:}} For the general model given in (\ref{55}), we have considered the parameter bounds as,
\begin{equation}\label{a7}
	H_0 \in [60,80] : \tilde{\lambda}\in\left(-0.5,-3/8\right] : \Delta \in [0,1]  :  \Omega^0_{m}\in[0,1] : \omega \in \left[0.5,1\right] : \alpha \in [0,0.5]
\end{equation}
Here, we have introduced two dummy parameters for the purpose of implementing the inequality constrains that where developed based on NEC and CEC requirements. These are given as,
\begin{equation}\label{a8}
	\Delta=\frac{3\zeta_1}{\omega} + \frac{\zeta_{0}\left[1+\tilde{\lambda}\left(3-\omega+3\zeta_{1}\right)\right]}{1-\tilde{\lambda}\zeta_{0}-\Omega^0_{m}}
\end{equation}
\begin{equation}\label{a9}
	\alpha=3\zeta_1+\omega
\end{equation}
From (\ref{a7}) one may notice that we have chosen $\omega \in [0.5,1]$ and have not considered $\omega<0.5$ this is because, in such cases, NEC and CEC were not simultaneously satisfied throughout the evolution.

\paragraph*{\textbf{Case Ia:}} For estimating best fit values of parameters associated with the model (\ref{60}), i.e., case Ia, we will assume the following priors for model parameters,
\begin{equation}\label{a10}
	H_0 \in [60,80] : \tilde{\lambda}\in\left(-0.5,-3/8\right]  : \Delta \in [0,1] :\Omega^0_{m}\in[0,1] : \omega \in \left[0.5,1\right] : \alpha \in [0,0.5]
\end{equation}
In this special case dummy variables defined in (\ref{a8}) and (\ref{a9}) changes as,
\begin{equation}\label{a11}
	\Delta=\frac{3\zeta_1}{\omega}
\end{equation}
\begin{equation}\label{a12}
	\alpha=3\zeta_1+\omega
\end{equation}

\begin{figure}
	\centering
	\includegraphics[width=0.8\columnwidth]{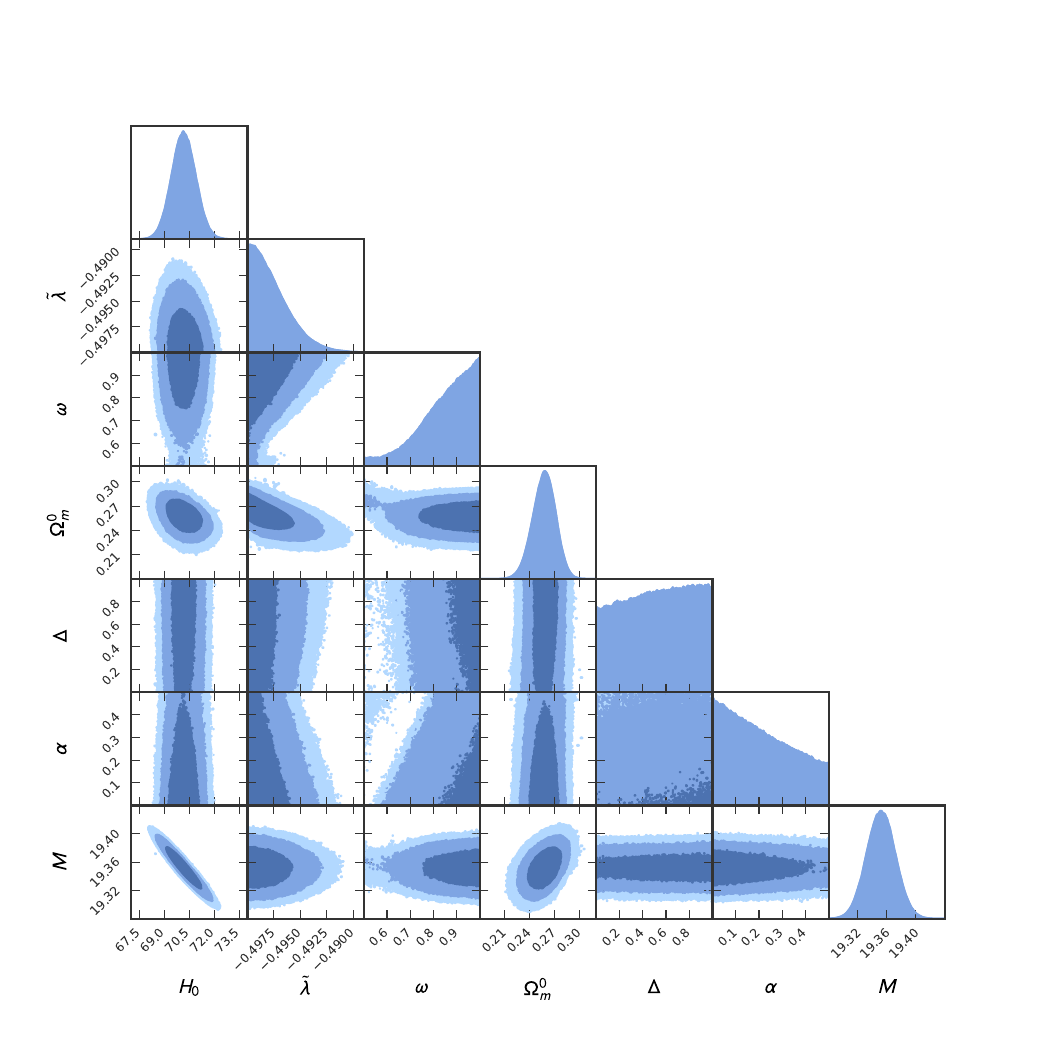}
	\caption{ Corner plot of 2D posterior contours with 1 sigma (68\%), two sigma (95\%) and three sigma confidence level (99.7\%) and 1D marginalized posterior distributions of model parameters for the combined OHD+SNe Ia data plotted using \cite{Bocquet2016} for the general case (i.e, Case I).}
	\label{CC1}
\end{figure}

\paragraph*{\textbf{Case Ib:}} For this special case we consider only the CEC and SLT to be the necessary requirement for the fluid and postulate that Eckart theory remains validity in the domain where fluid is far from equilibrium. This is because, during data analysis, when this special case was analyzed by applying NEC, the energy density of the vDM component became negative in certain epochs of evolution. Since we consider violation of CEC (having a negative energy density for vDM component) as unacceptable behavior for any model, we had to neglect such a scenario from analysis. However, it was observed that when NEC constraints are lifted, and only CEC and SLT are applied, model shows acceptable behavior. Hence, priors chosen in this case is,
\begin{equation}\label{a13}
	H_0 \in [60,80] :  \zeta_{1} \in [-1,1] : \zeta_{0}\in[-1,1] : \Omega^0_{m}\in[0,1] : \omega \in [0,1]  :  \tilde{\lambda}=\epsilon
\end{equation}
The best estimated value of model parameters associated with general model and its special cases, extracted by following $\chi^2$ minimization technique are provided in Table \ref{tab1}. Also, the a contour plot with one, two and three sigma confidence level for case I is provided as Fig. \ref{CC1}. 

\section*{Acknowledgments}
Vishnu A Pai acknowledges Cochin University of Science and Technology, Kochi, for providing financial support.

%\bibliography{ref.bib}
%\bibliographystyle{JHEP.bst}

\providecommand{\href}[2]{#2}\begingroup\raggedright\endgroup

\end{document}